\documentclass[conference,10pt]{IEEEtran}

\AtBeginDocument{%
  }

\usepackage{url}
\usepackage{booktabs}
\usepackage{graphicx}
\usepackage{tikz}
 
\usepackage{multirow}
\usepackage{listings}
\usepackage{tcolorbox}
\usepackage{color}
\usepackage{xcolor}
\usepackage{amsthm}
\usepackage{mdframed}

\newtcolorbox{findingsbox}[2][]{%
	colback=blue!5!white,
	colframe=blue!15,
	coltitle=black,
	fonttitle=\bfseries,
	title=#2,
	#1,
	boxrule=0.8pt,
	arc=4pt,
	left=6pt,
	right=6pt,
	top=6pt,
	bottom=6pt,
}

\usepackage{listings}
\usepackage{subcaption}
\usepackage{amsmath}
\usepackage{caption}

\lstdefinelanguage{json}{
	basicstyle=\small\ttfamily,
	numbers=none,
	breaklines=true,
	showstringspaces=false,
	literate=
	{"}{{{\color{green}"}}}1
	{:}{{{\color{blue}:}}}1
	{,}{{{\color{blue},}}}1
	{\{}{{{\color{blue}\{}}}1
	{\}}{{{\color{blue}\}}}}1
	{[}{{{\color{blue}[}}}1
	{]}{{{\color{blue}]}}}1,
	keywordstyle=\color{green}\bfseries,
}

\lstdefinestyle{jsonstyle}{
	language=json,
	basicstyle=\small\ttfamily,
	stringstyle=\color{blue},
	commentstyle=\color{green!50!black},
	frame=single,
	backgroundcolor=\color{gray!10},
	breaklines=true,
}

\theoremstyle{definition}        %
\newtheorem{definition}{Definition}

\newif\ifshowcomments
\showcommentstrue  %

\newcommand{\ra}{Repair\-Agent}
\newcommand{\oh}{Open\-Hands}
\newcommand{\acr}{Auto\-Code\-Rover}

\begin{document}

\author{
	\IEEEauthorblockN{Islem Bouzenia}
	\IEEEauthorblockA{
		CISPA Helmholtz Center for Information Security\\
		Germany\\
		bouzenia.islem@pm.me
	}
	\and
	\IEEEauthorblockN{Michael Pradel}
	\IEEEauthorblockA{
		CISPA Helmholtz Center for Information Security\\
		Germany\\
		michael@binaervarianz.de
	}
	\thanks{Both authors contributed equally to this work.}
}

\title{Understanding Software Engineering Agents:\\ A Study of Thought-Action-Result Trajectories}

\maketitle

\pagestyle{plain} %

\begin{abstract}
Large Language Model (LLM)-based agents are increasingly employed to automate complex software engineering tasks, such as program repair and issue resolution. These agents operate by autonomously generating natural language thoughts, invoking external tools, and iteratively refining their solutions. Despite their widespread adoption, the internal decision-making processes of these agents remain largely unexplored, limiting our understanding of their operational dynamics and failure modes. In this paper, we present a large-scale empirical study of the thought-action-result trajectories of three state-of-the-art LLM-based agents: \ra{}, \acr{}, and \oh{}. We unify their interaction logs into a common format, capturing 120 trajectories and 2,822 LLM interactions focused on program repair and issue resolution. Our study combines quantitative analyses of structural properties, action patterns, and token usage with qualitative assessments of reasoning coherence and feedback integration. We identify key trajectory characteristics, such as iteration counts and token consumption, recurring action sequences, and the semantic coherence of thoughts, actions, and their results. Our findings reveal behavioral motifs and anti-patterns that distinguish successful from failed executions, providing actionable insights for improving agent design, including prompting strategies, failure diagnosis, and anti-pattern detection. We release our dataset and annotation framework to support further research on transparent and robust autonomous software engineering agents.
\end{abstract}

\section{Introduction}
\label{sec:intro}

LLMs are becoming increasingly popular to automate various software engineering tasks.
Successful applications of LLMs include code completion~\cite{Chen2021,ziegler2022productivity,arXiv2024_De-Hallucinator,Barke2023}, automated program repair~\cite{Jiang2023,Xia2024a}, test case generation~\cite{Lemieux2023,Ryan2024,Yuan2024}, test oracle generation~\cite{Hayet2024a,Hossain2025}, and fuzz testing~\cite{icse2024-Fuzz4All}.
These approaches typically operate by prompting an LLM with a query, either a single time or in a hard-coded feedback loop.

Beyond querying LLMs with fixed prompt templates and within hard-coded algorithms, \emph{LLM agents} have recently emerged as another promising approach to software engineering automation.
Such agents are autonomous systems that use LLMs to iteratively reason about a problem, invoke existing tools, and adapt based on outputs produced by these tools~\cite{AgenticAISE2025}.
For example, LLM agents can autonomously fix bugs and other issues~\cite{icse2025-RepairAgent,zhang2024autocoderover,wang2024openhands,Yang2024a}, write tests to reproduce issues reported in natural language~\cite{Muendler2024,Ahmed2024,Issue2Test_arXiv2025}, and automate complex project setups~\cite{hu2025llm, issta2025_ExecutionAgent}.
To address a given task, an LLM agent autonomously plans and executes a sequence of actions, leveraging its reasoning capabilities and external tools.
Agents in software engineering typically use tools to interact with the code base and the environment, such as code editors, compilers, debuggers, and test runners.

While the success of LLM agents in software engineering is promising, the underlying decision-making processes remain largely opaque.
However, as their adoption increases, a critical question emerges:
\emph{How do LLM agents reach their solutions, and why do they (not) work well on specific tasks?}
Answering this question is crucial for understanding the strengths and limitations of LLM agents, as well as for improving their performance and reliability.
In particular, insights into the usage of tools could help to improve the design of LLM agents and their integration with existing software engineering tools.
Likewise, insights into the reasoning processes of agents could reveal common pitfalls and best practices, which will provide a basis for improving future LLM agents.

Understanding how LLM agents address software engineering tasks is important but also challenging.
On the upside, LLM agents provide detailed logs that document their reasoning and the actions they take.
We call these logs \emph{agent trajectories}, which we define as a sequence of thought-action-result triples, where each triple consists of a reasoning step, or thought expressed by an LLM, an external action (e.g., invoking a compiler), and the result of that action (e.g., the compiler output).
While the availability of agent trajectories is a promising starting point, there currently is no established methodology to systematically study them.

This paper presents the first in-depth empirical study of thought-action-result trajectories of LLM-based software engineering agents.
We focus on agents for automated program repair and issue solving, as this problem has received significant attention recently, and study trajectories from three state-of-the-art agents, namely \ra{}~\cite{icse2025-RepairAgent}, \acr{}~\cite{zhang2024autocoderover}, and \oh{}~\cite{wang2024openhands}.
Our study addresses three research questions:
\begin{itemize}
	\item RQ1: What are the overall properties of trajectories, such as the number of iterations and consumed LLM tokens?
	\item RQ2: What kinds of actions do agents perform, and what sequential patterns emerge within trajectories?
	\item RQ3: How do agents develop and adjust their reasoning and actions throughout task execution, and how coherent are thoughts, actions, and feedback?
\end{itemize}
As a cross-cutting theme, we also investigate how the answers to the above questions differ for trajectories that successfully complete a task versus those that fail.

To address these questions, we introduce a novel methodology that analyzes LLM-based software engineering agents through their thought-action-result trajectories.
We begin by collecting logs from the three studied agents (RepairAgent, AutoCodeRover, OpenHands) and convert them into a unified trajectory format.
We then compute statistics to examine trajectory properties and how they relate to success or failure of an agent (RQ1).
Next, we manually categorize agent actions and mine frequent action sequences (RQ2).
Finally, to study reasoning coherence (RQ3), we apply open coding to label semantic relationships between thoughts, actions, and results.
This mixed-method approach enables us to extract both quantitative insights and qualitative interpretations of agent behavior. Our study leads to several key findings:

\begin{itemize}
	\item Test-driven repair agents (\ra{}, \oh{}) have longer, token-intensive failing trajectories, reflecting task complexity; conversely, \acr{}'s streamlined retrieve-locate-fix workflow is more efficient.
	
	\item Successful action sequences balance exploration, explanation, fix generation, and testing. In contrast, failures exhibit repetitive, non-adaptive cycles, such as repeated identical actions without follow-up, and debugging anti-patterns, such as generating a fix without testing it.
	
	\item Semantic alignment between thoughts and actions is essential: Even rare misalignments strongly correlate with failure or increased computational cost, underscoring the need for explicit validation mechanisms (e.g., self-reflection or critique) to ensure coherence.

	\item Agents differ in reasoning dynamics: Flexible state-machine agents enable revisiting and steady progress; test-driven agents prioritize systematic verification; streamlined agents focus on concise retrieve-and-fix workflows. Identifying the optimal architecture for each task warrants further research.

\end{itemize}

Our work relates to prior efforts in LLM interpretability, log analysis, and trajectory-based agent improvement.
LLM interpretability techniques provide insights into why a model produces a specific output~\cite{Guidotti2019,survey_explain_llm}, e.g., via counterfactual reasoning~\cite{Cito2022} or by highlighting specific code locations~\cite{ying2019gnnexplainer}.
Unlike these approaches, our work focuses on the trajectory of an agent's reasoning and actions over time, rather than just a single prediction.
To help users inspect agent behavior, prior work has proposed tools for visualizing the interactions within multi-agent systems~\cite{lu2024agentlens,Epperson2025}.
Our work complements these efforts by focusing on a single agent that interacts with its environment, and by targeting agents in software engineering.
Finally, several techniques aim at fully automatically improving agents via reinforcement learning based on past trajectories~\cite{Gupta2024,song2024trial,deng2024novice}.
Instead, our work provides insights that enable agent developers to understand and improve agents.

In summary, this paper contributes the following:
\begin{enumerate}
	\item \emph{Methodology}. We introduce a novel trajectory-based methodology to study the interplay of thoughts, actions, and results in software engineering agents.
	\item \emph{Empirical results}. We provide the first systematic study of agent trajectories across multiple state-of-the-art software engineering agents, revealing characteristic patterns and challenges.
	\item \emph{Insights and implications}. We offer insights into agent decision-making processes that inform better prompting strategies, trajectory supervision, and evaluation metrics.
\end{enumerate}

\section{Methodology}

\begin{figure*}
	\centering
	\includegraphics[width=.7\linewidth]{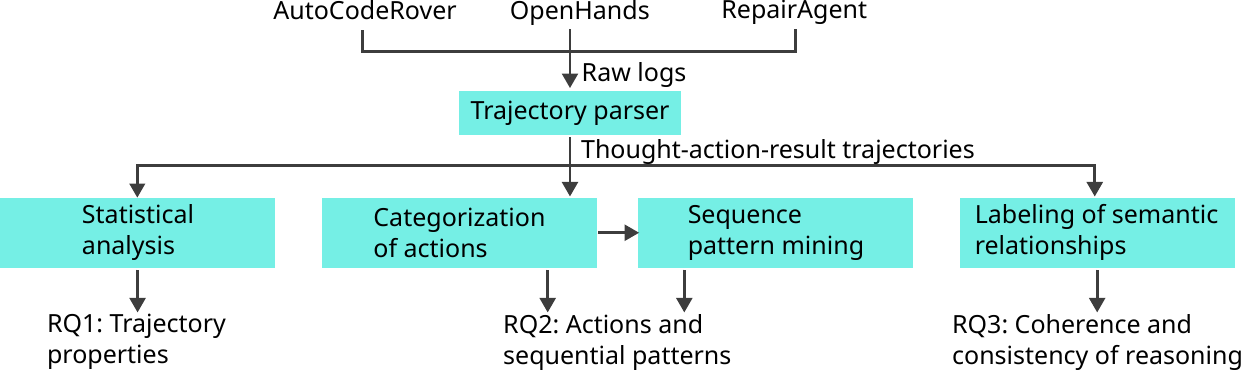}
	\caption{Overview of our methodology.}
	\label{fig:methodologyoverview}
\end{figure*}

To answer the research questions given in Section~\ref{sec:intro}, we present a novel methodology for analyzing the trajectories of LLM agents for software engineering.
Figure~\ref{fig:methodologyoverview} provides an overview of our methodology, which consists of five main components.
Given the raw logs of an LLM agent, we first parse the logs into a unified representation of trajectories.
We then address RQ1 by computing statistics, e.g., about the length of trajectories and the costs imposed by the agents.
To study the actions performed by the agents (RQ2), we systematically inspect and categorize the actions, followed by mining the resulting action sequences for sequential patterns.
Finally, we analyze the semantic relationships between thoughts, actions, and result (RQ3) through open coding, allowing us to study the coherence and consistency of an agent's behavior.
The remainder of this section describes our data collection and each of the components of our methodology in detail.

\subsection{Data Collection}
We collect trajectory logs from three state-of-the-art software engineering agents:
\begin{itemize}
	\item \emph{\ra{}\footnote{https://github.com/sola-st/RepairAgent}~\cite{icse2025-RepairAgent}:} An autonomous agent for program repair evaluated on 835 bugs of Defects4J~\cite{Just2014}. We consider the main trajectories of \ra{} and we ignore the mutation sub-trajectory
	\item \emph{\acr{}\footnote{https://github.com/AutoCodeRoverSG/auto-code-rover}~\cite{zhang2024autocoderover}:} An agent designed for automated issue resolution, evaluated on  the 300 instances of SWE-bench Lite~\cite{Jimenez2023}. We utilize the trajectories from \acr{} V1. Although subsequent versions of \acr{} exist, their trajectories were unavailable to us. Additionally, \acr{} conducts three runs and reports average results; since there is no fundamental difference among these runs, we selected the trajectories from the first run for this study.

	\item \emph{\oh{}\footnote{https://github.com/All-Hands-AI/OpenHands}~\cite{wang2024openhands}:} A general-purpose code editing agent, also evaluated on the 300 tasks in SWE-bench Lite~\cite{Jimenez2023}. We use trajectories obtained from OpenHands CodeAct, a version that is tailored for issue resolution.
\end{itemize}

To ensure a diverse yet manageable dataset for detailed manual annotation, we randomly sample 40 trajectories per agent. Given the average success rate of approximately 27\% across the three agents, we proportionally sample about 10 successful trajectories per agent. Each trajectory captures a complete bug repair or issue resolution through iterative agent-environment interactions. Our dataset thus comprises 120 trajectories totaling 2,822 iterations across all agents.

\subsection{Trajectory Parsing and Representation}

To ensure consistency in our analysis, all selected trajectories are processed using our \emph{trajectory parser}.
which converts the raw logs from different agents into a unified representation:

\begin{definition}
\label{def:agent_trajectory}
An \emph{agent trajectory} is a sequence of \emph{iterations} that each consist of three components:
$$\mathcal{T} = [(t_1, a_1, r_1), (t_2, a_2, r_2), \dots, (t_n, a_n, r_n)]$$
where each tuple \((t_i, a_i, r_i)\) represents the following:
\begin{itemize}
	\item The \emph{thought} \(t_i\) is a natural language description of the agent's internal reasoning, such as a diagnostic insight, an explanation of program behavior, or a proposed fix.
	\item The \emph{action} \(a_i\) is an invocation of an external operation, such as editing code, running a test, or another tool.
	\item The \emph{result} \(r_i\) is the response to action \(a_i\), such as compiler output, test results, execution traces, or error logs.
\end{itemize}
\end{definition}

To convert raw logs into a structured format, we implement parsers for each agent. \ra{} and \oh{} produce structured JSON logs with clearly separated reasoning, actions, and results, each requiring tailored parsing due to distinct JSON schemas. For \acr{}, we develop a heuristic to map its semi-structured output into trajectories by detecting tool calls via tool names, extracting preceding reasoning as thoughts, and capturing the call results as separate text items.

\subsection{Statistical Analysis}

To address RQ1, we compute several trajectory-level metrics that capture the overall properties of agent trajectories.
Given a trajectory $\mathcal{T} = [(t_1, a_1, r_1), \dots, (t_n, a_n, r_n)]$, we compute the following metrics.

\begin{definition}
	The \emph{trajectory length} of $\mathcal{T}$ is the number $|\mathcal{T}|$ of iterations in the trajectory.
\end{definition}

The trajectory length may be affected by various factors, including task complexity, effort spent by an agent, and the verbosity of the agent's reasoning.
To manage computational effort, agents limit the trajectory length, e.g., \ra{} caps at 40 iterations, while \oh{} allows up to 100 iterations.

\begin{definition}
The \emph{trajectory cost} of $\mathcal{T}$ is the total number of tokens consumed by the agent during the trajectory:
$$\sum_{i=1}^{n} \left( \mathit{tokens}(t_i) + \mathit{tokens}(a_i) + \mathit{tokens}(input_i) \right)$$
where $\mathit{tokens}()$ denotes the number of tokens consumed by or generated by the LLM and $\mathit{input_i}$ denotes the input prompt at iteration $\mathit{i}$, which integrates $\mathit{r_i}$.
\end{definition}

The trajectory cost is important because queries to LLMs are typically charged based on the number of tokens processed.

\begin{definition}
The \emph{success} of a trajectory $\mathcal{T}$ is a binary label indicating the agent's success in completing the task.	
\end{definition}

We define success as reported in the original evaluation of the agents, i.e., whether the agent produced a valid patch or solution to the task.
Specifically, for \acr{} and \oh{}, we consider a trajectory successful if the agent produces a patch that passes the tests executed by SWE-bench~\cite{Jimenez2023}.
For \ra{}, we consider a trajectory successful if the produced patch passes the tests executed by Defects4J~\cite{Just2014} (called ``plausible'') and manual inspection confirms that the patch is equal or semantically to the developer-produced patch (called ``correct'').

\subsection{Categorizing Actions}

To address RQ2, which investigates the actions performed by agents, we categorize the actions in the trajectories.
A key challenge is the diversity of actions performed by different agents, which includes various operations, such as code editing and executing tests, while not following a unified naming convention to refer to actions.
After an initial inspection of trajectories, we observe that the actions performed by agents can be grouped into a small number of high-level categories that reflect activities developers commonly perform when debugging or fixing code~\cite{zeller2009programs}.

Based on this observation, we categorize each action into one of the following eight categories:

\begin{enumerate}
    \item \emph{Explore}: Listing, navigating, and reading files to gather contextual information.
	\item \emph{Locate}: Identifying specific code regions or elements relevant to the bug or issue.
	\item \emph{Search}: Employing search tools to find occurrences of code elements, symbols, or patterns expressed via regular expressions.
    \item \emph{Reproduce}: Creating test cases to confirm the presence of the bug or the issue.
 	\item \emph{Generate fix}: Proposing or writing code modifications intended to resolve the issue.
    \item \emph{Run tests}: Executing test cases to validate the proposed fix or to validate the test cases themselves.
    \item \emph{Refactor}: Improving code quality through optimization, reformatting, or commenting.
    \item \emph{Explain}: Producing natural language explanations or summaries of the issue or applied fix.
\end{enumerate}

To categorize the actions, we first map known tools used by agents, such as search tools into the respective category (e.g., Search). Then, we manually inspect the remaining actions that we were unable to map automatically (e.g., a terminal command).
The labeling is initially performed by one author, followed by a discussion of ambiguous and uncertain cases to resolve them collaboratively among both authors.
A small subset of actions (8.3\%) does not fit into any of the above categories, such as internal state transitions in \ra{} and LLM-suggested actions that \acr{} fails to parse.
We exclude these cases from subsequent analyses.

\subsection{Mining Sequential Action Patterns}

We analyze sequences of categorized actions in agent trajectories to identify recurrent decision-making patterns. Each trajectory is represented as a sequence of action labels, e.g., \textit{Explore} \(\rightarrow\) \textit{Locate} \(\rightarrow\) \textit{Generate Fix} \(\rightarrow\) \textit{Reproduce} \(\rightarrow\) \textit{Run Tests}.
We study these sequences by mining for frequently occurring subsequences. First, we implement a lightweight Python script to extract fixed-length $n$-grams and compute their frequencies, choosing $n$ large enough to avoid trivial short sequences, but small enough to retain sufficient diversity and easy-to-analyze patterns. From our experiments, $n \in \{4,5,6\}$ performs well, and for this study we set $n = 4$. Next, for each agent we identify the most frequent 4-grams and analyze how their relative frequencies differ between successful and failed tasks. Finally, we flag patterns overrepresented in failing trajectories, as well as rare or unusual 4-grams that may indicate nonstandard or suboptimal workflows.

\subsection{Labeling Semantic Relationships}
\label{sec:semantic_labeling}

\begin{table*}[t]
	\centering
	\setlength{\tabcolsep}{3pt}
	\caption{Semantic relationships between parts of agent trajectories.}
	\begin{tabular}{@{}lp{16cm}@{}}
		\toprule
		Relationship & Definition and example \\
		
		\midrule
		\multicolumn{2}{@{}l@{}}{\emph{Thought $t_i$ $\rightarrow$ action $t_i$ (within one iteration):}} \\
		\midrule
		Alignment & The action logically aligns with the thought. Example: Thought ``I need to inspect usages of function foo'' aligns with action ``search\_calls('foo')''. \\
		Misalignment & The action does not reflect the thought. Example: Thought ``I need to write a fix for this bug'' misaligns with action ``read documentation''. \\
		
		\midrule
		\multicolumn{2}{@{}l@{}}{\emph{Thought $t_i$ $\rightarrow$ thought $t_{i+1}$ (across consecutive iterations):}} \\
		\midrule
		Follow-up & The thought is a logical, intuitive follow-up to the previous one.  Example: ``Line 161 indicates a potential off-by-one error'' followed by ``I should verify and adjust the loop boundary at line 161''. \\
		Refinement & The thought adds specific details to the previous thought. Example: ``I should focus on collecting information to fix the bug based on the expected and actual outputs in the failing test case'' refines ``I should focus on the assertion at line 342 in ExtensionFunctionTest.java involving comparing the expected and actual output values''. \\
		Redundancy & The thought is repeated without added new insights. We count both semantic and syntactic repetition. Example: ``The bug is in function A'' is redundant with ``The bug seems to happen in function A.'' \\
		Divergence & The thought shifts focus to a different direction. Example: ``I will now suggest a fix...'' diverges from ``I need to find the location of the bug.''\\
		Contradiction  & The thought contradicts the previous thought. Example: ``I think the error is in function A'' contradicts ``The error is actually in function B, as shown by the stack trace''. \\		
		
		\midrule
		\multicolumn{2}{@{}l@{}}{\emph{Action $a_i$ $\rightarrow$ action $a_{i+1}$ (across consecutive iterations):}} \\
		\midrule
		Follow-up  & The action is a plausible follow-up to the previous action. Example: Running test cases after writing a fix.\\
		Refinement  & The action improves or extends the previous action. Example: ``Search for 'error handling'{}'' is refined to ``Search for 'error handling in function X'{}''. \\
		Repetition & The same action is taken again without modification with the same parameters (syntactically or semantically). Example: Running the same test case multiple times without changing the code. \\
		Divergence & The action does not follow naturally from the previous one. Example: ``Trying a candidate patch'' followed by ``Making search queries''. \\

		\midrule
		\multicolumn{2}{@{}l@{}}{\emph{Result $r_i$ $\rightarrow$ thought $t_{i+1}$ (across consecutive iterations):}} \\
		\midrule
		Follow-up & The thought is a logical follow-up on the results of the previous iteration. Example: Result ``Test passed'' followed by thought ``Since test cases passed, I think I can end my task''. \\
		Refinement & The result provides specific details that refine the next thought. Example: Detailed compiler errors refine the hypothesis about the bug location. \\
		No influence & The result does not affect the next thought. Example: Result ``New search results: ...'' followed by thought ``I need to suggest a fix''.\\
		Misinterpretation & The thought misinterprets the previous result. Example: An error output followed by thought ``The bug was fixed''. \\
		
		\midrule
		\multicolumn{2}{@{}l@{}}{\emph{Result $r_i$ $\rightarrow$ action $a_{i+1}$ (across consecutive iterations):}} \\
		\midrule
		Informative & The result provides information that is incorporated into the next action. Example: A specific error message is used as an argument in the next code search. \\
		Triggering & The result directly triggers a particular action. Example: Successful bug fixing triggers the end of task. \\
		No influence & The result has little or no effect on the next action. 
		Example: The LLM obtains new search results but none of them affect the fix suggested next. \\
		
		\bottomrule
	\end{tabular}
	
	\label{tab:relationships}
\end{table*}

RQ3 investigates the semantic relationships among trajectory components (thoughts, actions, results) to evaluate the coherence and consistency of an agent's behavior. We systematically label relationships both within iterations (e.g., thought to action) and across iterations (e.g., how a result in iteration $i$ influences the thought in iteration $i+1$).

To label relationships, we apply an open coding approach~\cite{corbin1990basics} to the trajectories.
Open coding is a qualitative analysis technique where a researcher identifies and labels recurring patterns within a dataset.
Beginning with no predefined labels, the coding evolves through manual inspection as new patterns emerge.
We apply this technique to five types of relationships between components of agent trajectories.
The result of this process is the set of relationships presented in Table~\ref{tab:relationships}.
The following summarizes the five types of relationships we consider, and examples of the relationships we identify.

\paragraph{Thought $t_i$\,$\rightarrow$\,action $a_i$}
These relationships capture the degree to which an action that the agent choses faithfully implements the immediately preceding thought expressed by the agent.
For example, we assign an \emph{Alignment} label when the action logically follows from the thought, such as thinking ``I need to inspect usages of \texttt{foo}'' followed by searching the code base for occurrences of \texttt{foo}.

\paragraph{Thought $t_i$\,$\rightarrow$\,thought $t_{i+1}$}
These relationships describe how the agent's internal reasoning evolves across iterations.
For example, we label a pair of thoughts as \emph{Follow-up} when the second thought is a natural continuation of the first, such as ``I need to inspect usages of \texttt{foo}'' followed by ``I should check the function definitions that use \texttt{foo}''.
In contrast, we label a pair of thoughts as \emph{Divergence} when the second thought pivots to a different line of reasoning, such as ``I need to inspect usages of \texttt{foo}'' followed by ``I will now suggest a fix for the bug''.

\paragraph{Action $a_i$\,$\rightarrow$\,action $a_{i+1}$}
These relationships characterize how the agent's operations progress toward task completion.
For example, we label a pair of actions as \emph{Repetition} when the agent performs the same action consecutively without new effect on the agent and codebase.

\paragraph{Result $r_i$\,$\rightarrow$\,thought $t_{i+1}$}
These relationships characterize how the agent incorporates feedback from its environment into the next thought.
For example, we label a pair of result and thought as \emph{Follow-up} when the thought is a logical continuation of the previous result, such as ``Tests passed'' followed by ``Since test cases passed, I think I can end my task''.
In contrast, we assign the label \emph{Misinterpretation} when the agent understands a result incorrectly, such as ``Test failed'' followed by ``The bug was fixed''.

\paragraph{Result $r_i$\,$\rightarrow$\,action $a_{i+1}$}
These relationships capture how the agent's results influence its subsequent actions.
For example, we label a pair of result and action as \emph{Informative} when the result provides information that is incorporated into the next action (e.g., an API name found through search).

We focus on these five specific types of relationships because they directly capture the immediate and most influential semantic connections within and across consecutive iterations. Long-term sequence relationships or more distant dependencies can be informative but not straightforward to systematically label and interpret due to their extension and complexity. Our targeted scope balances depth and feasibility.

The relationships in Table~\ref{tab:relationships} are the result of labeling 40 trajectories from each of the three agents, totaling 14K pairs of trajectory components.
The annotation process was initially conducted by the first author over five months of full-time work.
To ensure quality and consistency, both authors collaborated closely to discuss the set of labels, resolve ambiguous and difficult cases, and cross-validate individual annotations. Moreover, the main annotator periodically revisited previously labeled examples to ensure consistency across trajectories and re-investigated abnormal numbers (e.g., high misalignment labels in an agent).

\section{Results}
\label{ref:analysis}
	
This section presents a detailed analysis of the three research questions introduced earlier.

\subsection{\textbf{RQ1}: Trajectory-Level Properties}
We investigate trajectory properties aiming to understand how their iterative processes differ in length and complexity. We answer this question by quantitatively analyzing the distribution of iteration counts and token consumption.

\begin{figure}
	\centering
	\includegraphics[width=1\linewidth]{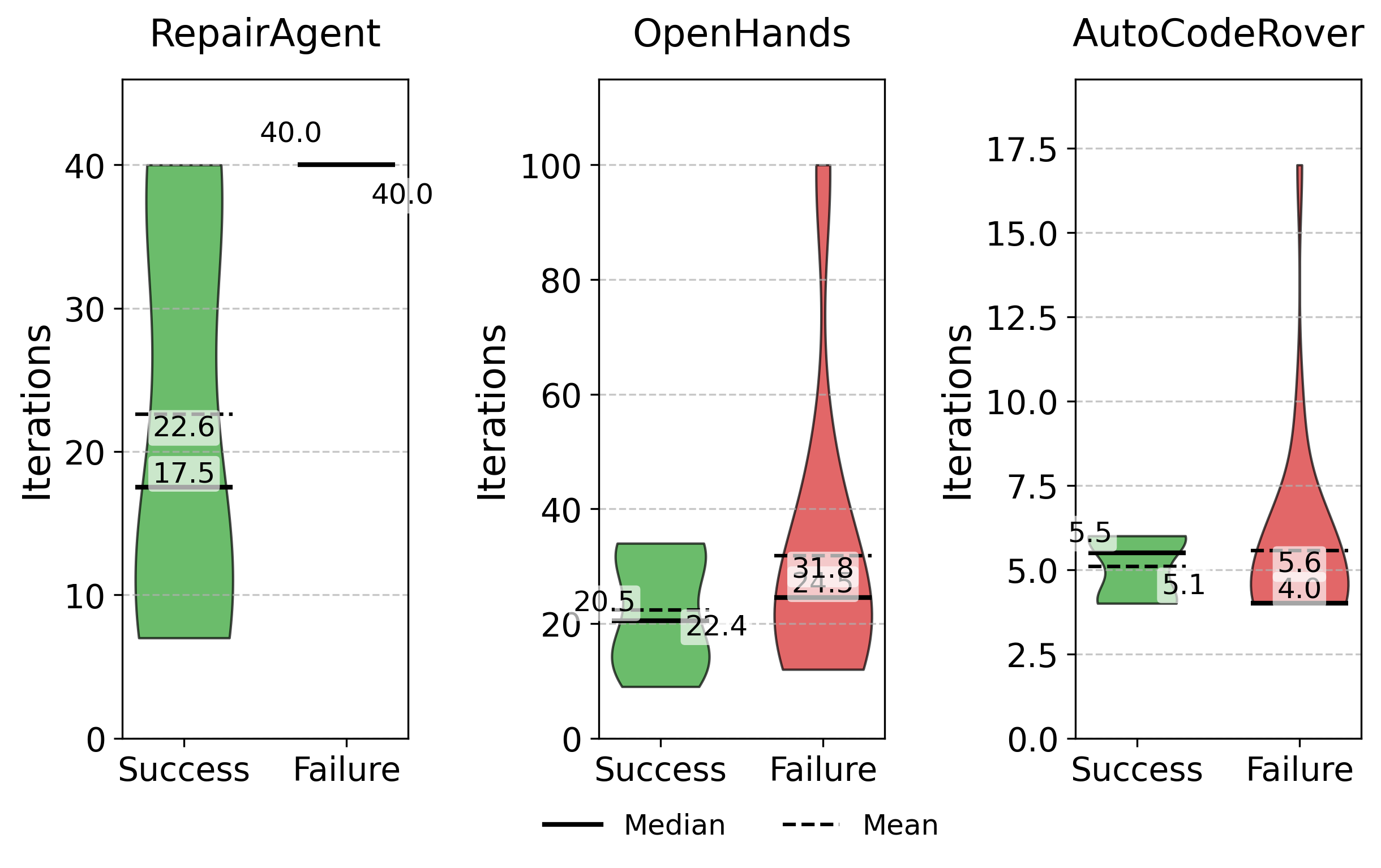}
	\caption{Comparative trajectory length between successful and unsuccessful trajectories. Note that the y-axis differs between the agents.}
	\label{fig:trajlengthsplit}
\end{figure}

Figure~\ref{fig:trajlengthsplit} shows the distribution of iteration counts across agents. \textit{\ra{}} has a mean trajectory length of 34 iterations, indicating consistent iterative reasoning. Similarly, \textit{\oh{}} averages 29 iterations, reflecting comparable depth. In contrast, \textit{\acr{}} (ACR) completes workflows in fewer steps (6 on average), highlighting a more efficient patch generation process.

The figure also compares trajectory lengths between successful and unsuccessful instances, revealing agent-specific patterns. For \ra{}, unsuccessful cases have longer trajectories (mean 40) than successful ones (mean 22), indicating higher effort on difficult tasks, consistent with its design to continue attempts until the agent finds a fix or exceeds the budget. \oh{} shows a similar trend, though a bit less pronounced, with some unsuccessful trajectories reaching the 100-iteration cap. For \acr{}, successful and unsuccessful trajectories have similar average lengths, but unsuccessful ones exhibit higher variability and outliers, often due to internal errors like \textit{Response Parse Error}. \acr{} terminates after patch application without integrating tests into its workflow.

\begin{findingsbox}{Discrepancies in Trajectory Length}
	Test-driven repair leads to longer trajectories, especially in unsuccessful cases, reflecting sustained exploration for solutions. In contrast, agents with direct patch application show shorter trajectories, with unsuccessful cases marked mainly by greater variance.
\end{findingsbox}

\begin{figure*}
	\centering
	\includegraphics[width=1\linewidth]{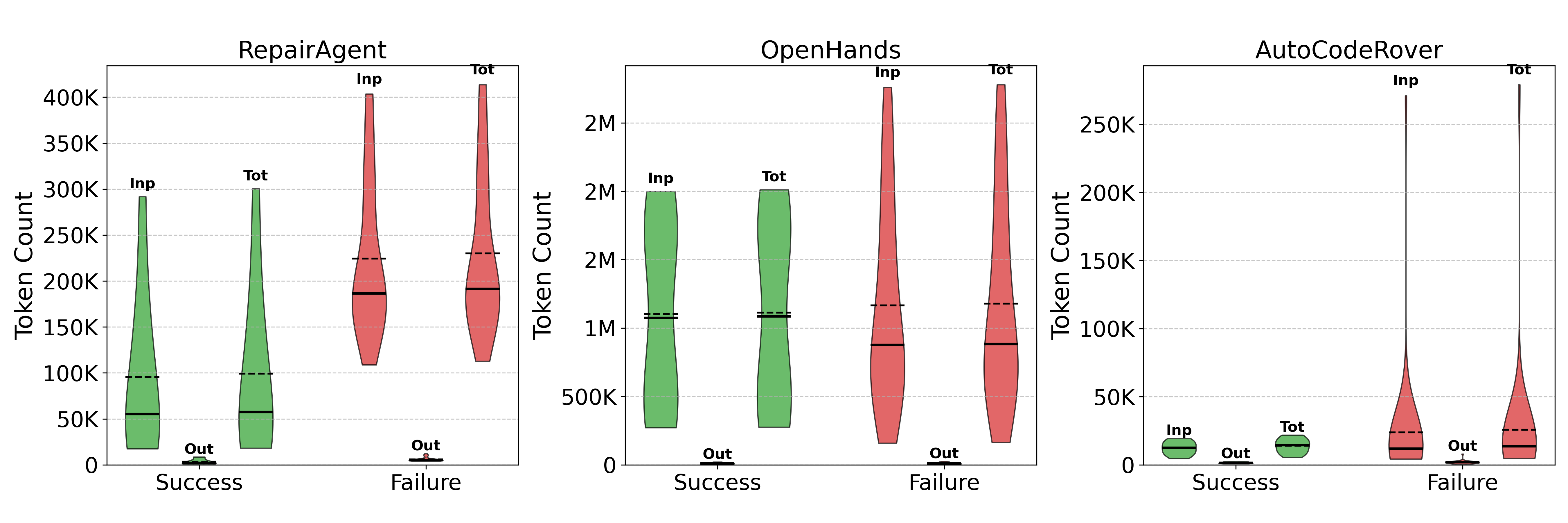}
	\caption{Input, Output, and Total tokens over successful trajectories and unsuccessful ones. The y-axis differs for each agent.}
	\label{fig:tokenusage}
\end{figure*}

Figure~\ref{fig:tokenusage} presents cumulative token consumption per trajectory.
\acr{} exhibits the smallest footprint (mean 23K, median 14K tokens).
\oh{} averages about 1.2M tokens (median 900K), approximately 52 times more than \acr{}.
\ra{} demonstrates moderate usage (mean 220K, median 180K).
Key observations are as follows:

\begin{itemize}
	\item \textbf{\ra{}}: Unsuccessful debugging attempts use more tokens than successful ones, reflecting extended iterations and additional context (e.g., search).
	\item \textbf{\oh{}}: Successful trajectories consume more input tokens than unsuccessful ones, while output token usage is similar (more input context used in successful runs).
	\item \textbf{\acr{} (ACR)}: Trajectories are consistently short and efficient. Failures often involve \textit{Error Response Parsing} sequences that modestly increase token usage.
\end{itemize}

\begin{findingsbox}{Token Consumption Patterns}
\ra{}'s token usage increases with trajectory length due to its iterative nature. \oh{}'s higher token usage in successful runs suggests that input context aids success. \acr{} maintains efficient consumption, with variability mainly from parsing errors.
\end{findingsbox}

These findings indicate that both iteration count and total token consumption provide complementary and insightful metrics for assessing problem complexity and the behavioral dynamics of each agent. Subsequent research questions dive deeper into the precise characterization of agents behavior.

\subsection{\textbf{RQ2}: Actions and Patterns of Actions Sequences}
\subsubsection{Overall Action Usage}

Figure~\ref{fig:actionsfrequency} compares the overall distribution of action categories.
Across all agents, the most frequent actions are \textit{Generate Fix} (23\%), \textit{Run tests} (19\%), \textit{Search} (15\%), and \textit{Explore} (14\%).
While these core actions are common, each agent exhibits distinct patterns in action selection and sequencing.

\ra{} tends to emphasize a balanced approach incorporating search and exploration, fix generation, and comprehensive testing. In contrast, \oh{} places particular emphasis on a test-driven methodology, including the generation of issue-reproducing tests, which is distinctive to \oh{}'s workflow. \acr{} operates through a structured search–locate–fix process, typically omitting explicit \textit{Reproduce} and \textit{Test} steps in its main pipeline. It is worth noting that a subsequent extension, SpecRover~\cite{ruan2024specrover}, introduces dedicated reproduction and testing phases.

\subsubsection{Action Usage Over Task Progress}

\begin{figure*}[!htbp]
	\centering
	\begin{subfigure}[b]{0.49\textwidth}
		\centering
		\includegraphics[width=\linewidth]{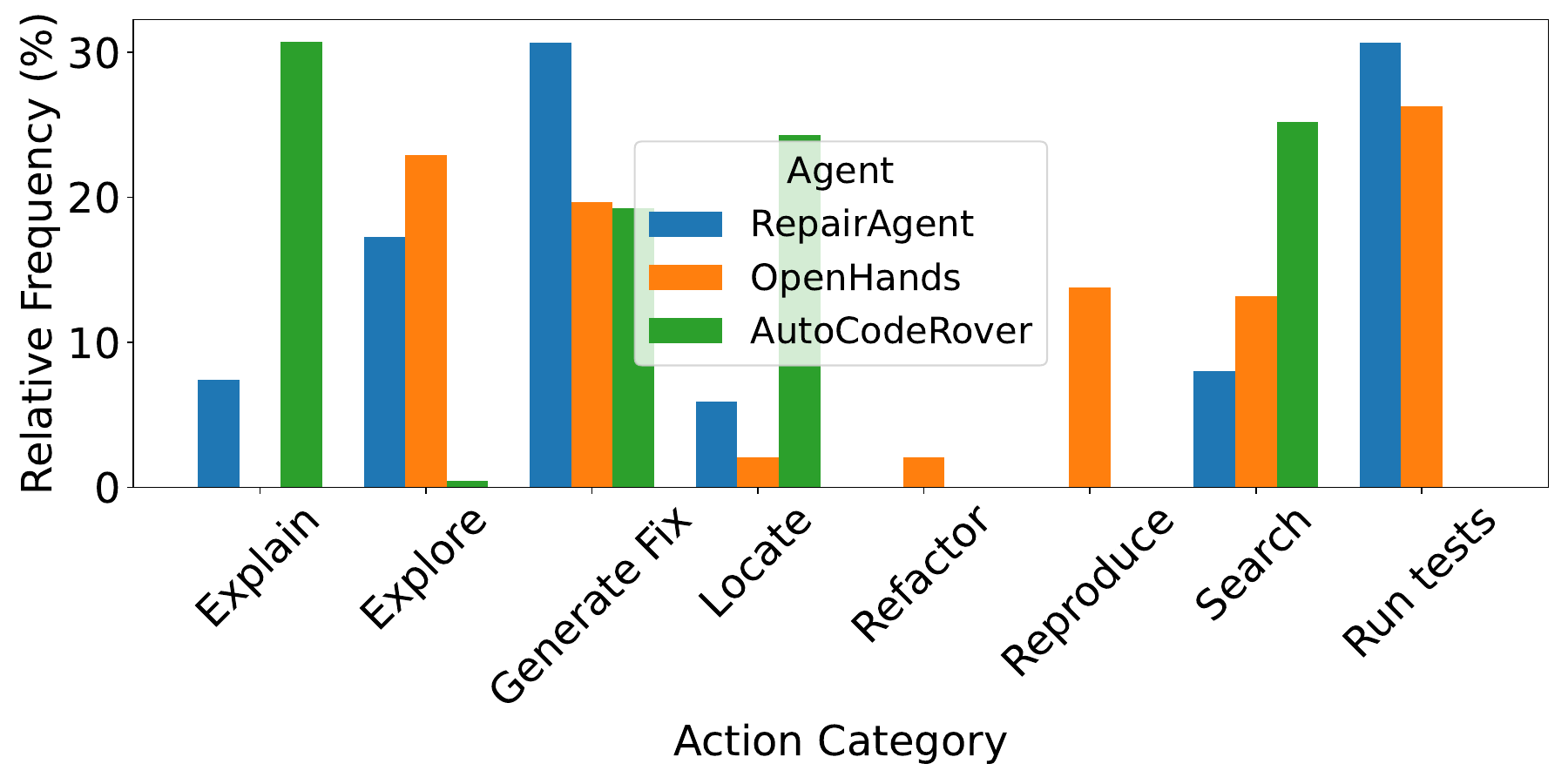}
		\caption{Actions frequency across agents.}
		\label{fig:actionsfrequency}
	\end{subfigure}
	\hfill
	\begin{subfigure}[b]{0.49\textwidth}
		\centering
		\includegraphics[width=\linewidth]{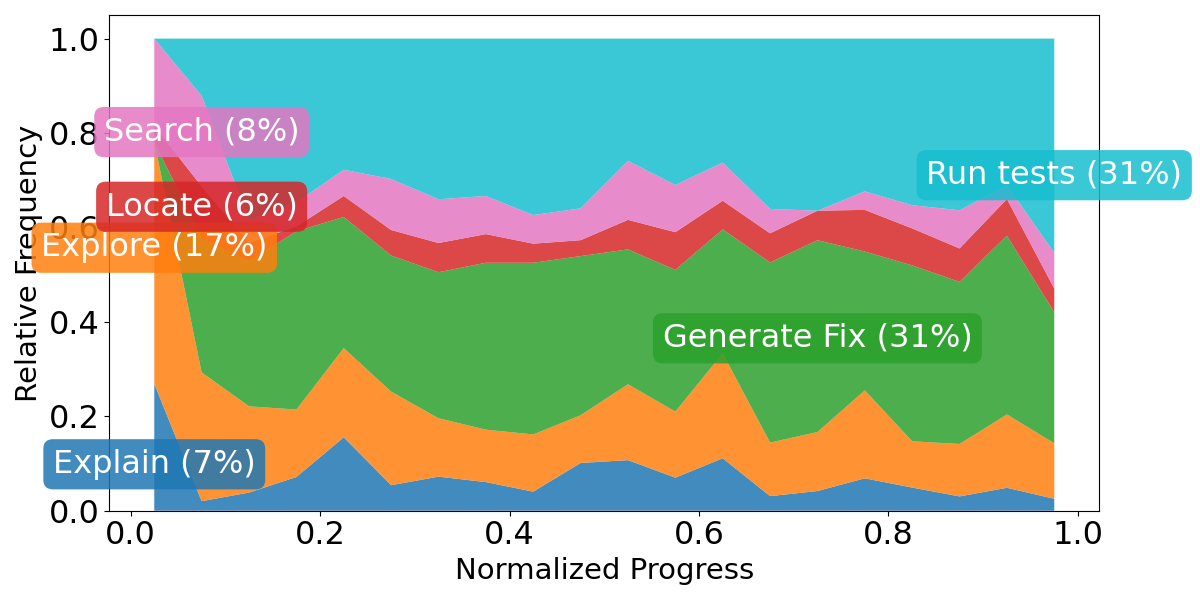}
		\caption{\ra{}}
		\label{fig:stackedrepairagent}
	\end{subfigure}
	
	\hfill
	
	\vspace{0.5cm} %
	
	\begin{subfigure}[b]{0.49\textwidth}
		\centering
		\includegraphics[width=\linewidth]{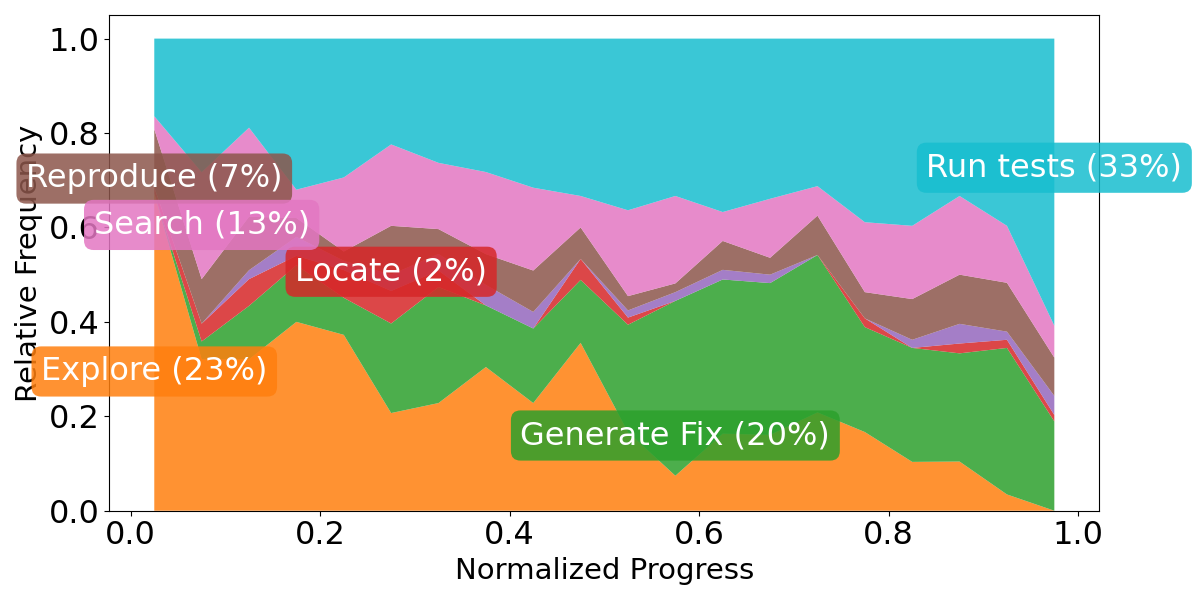}
		\caption{\oh{}}
		\label{fig:stackedcodeact}
	\end{subfigure}
	\begin{subfigure}[b]{0.49\textwidth}
		\centering
		\includegraphics[width=\linewidth]{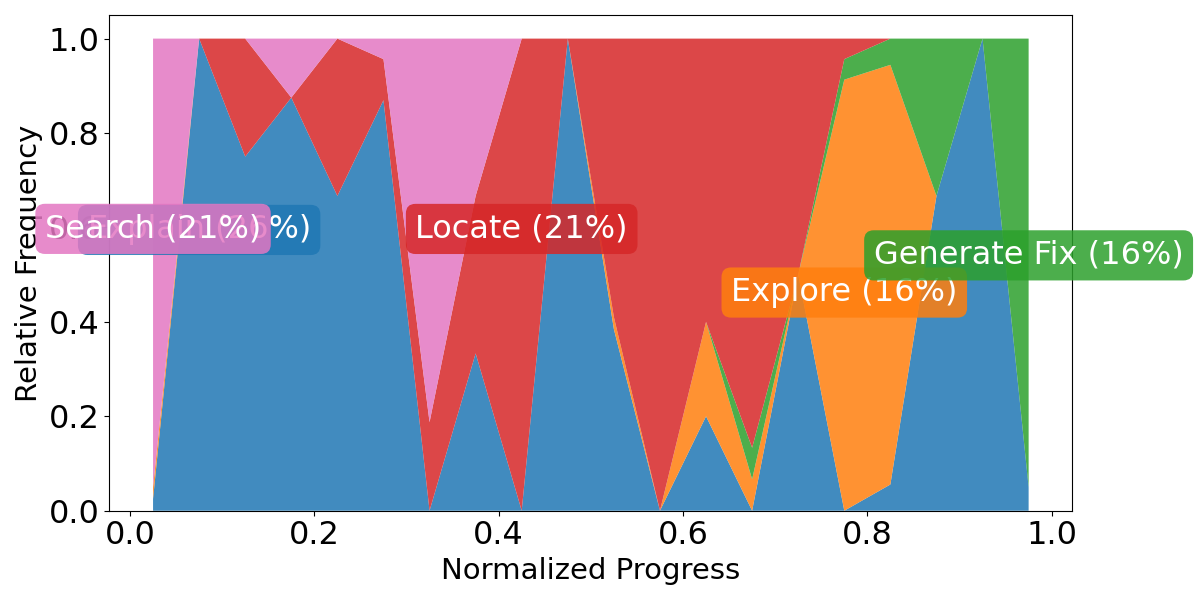}
		\caption{\acr{}}
		\label{fig:stackedacr}
	\end{subfigure}
	
	\caption{Normalized action distribution across task execution phases for each agent.}
	\label{fig:stacked_all_agents_grid}
\end{figure*}

To analyze behavior evolution, we normalize iterations \(i \in [1..N]\) of a trajectory to percentage progress. Figures~\ref{fig:stackedrepairagent}–\ref{fig:stackedacr} depict the proportion of actions per category at each progress point (e.g., 20\% of the total trajectory length). In \ra{} (Fig.~\ref{fig:stackedrepairagent}), the initial 0–20\% of the trajectory is evenly split between \textit{Fix Generation} plus \textit{Running Tests} (~50\%) and the remaining categories (\textit{Explore}, \textit{Explain}, \textit{Search}, \textit{Locate}). From 20\% to 100\%, \textit{Fix Generation} and \textit{Running Tests} dominate (~70\%). \oh{} (Fig.~\ref{fig:stackedcodeact}) starts with ~50\% \textit{Explore} and \textit{Search} in the first 20\%, then shifts towards increased \textit{Fix Generation}, \textit{Reproduce}, and \textit{Running Tests}, which constitute ~75\% in the last 20\%. \acr{} (Fig.~\ref{fig:stackedacr}), with median \(N=5\), spends the first 50\% on \textit{Search}, \textit{Locate}, and \textit{Explain}, then transitions to \textit{Locate} (30\%) and \textit{Generate Fix} (20\%) in the latter half.
The sharp transitions in \acr{}'s action distribution are due to its short trajectories.
Failing trajectories include many \textit{ErrorResponse} events, omitted from the plot.

\begin{findingsbox}{Takeaways from Agent Actions}
Developers can use actions distribution as a secondary metric in evaluating the quality of agents by analyzing the frequency and timestamps of actions or by comparing them to real trajectories obtained from human experts.
\end{findingsbox}

\subsubsection{Sequence Patterns}

\begin{figure}
	\centering
	\includegraphics[width=1\linewidth]{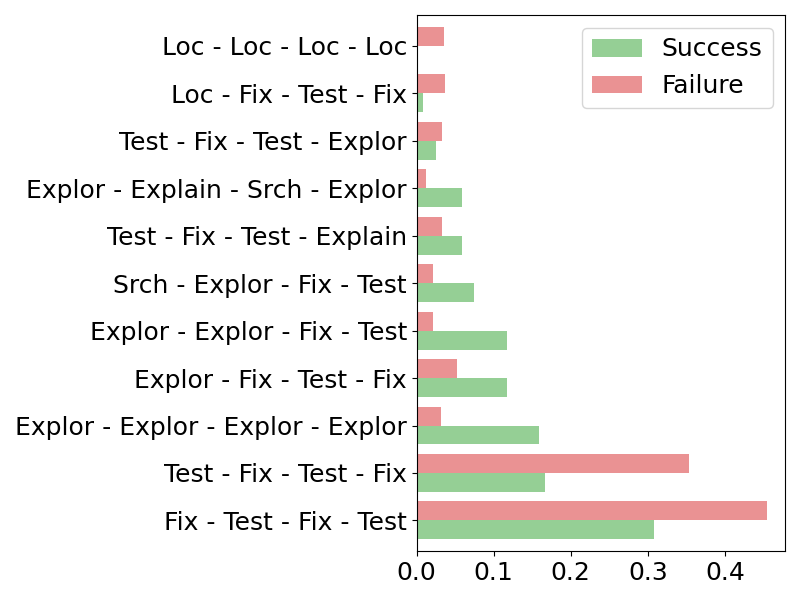}
	\caption{Top 11 sequences of action 4-grams (\ra{}).}
	\label{fig:seqrepairagentbars}
\end{figure}

\begin{figure}
	\centering
	\includegraphics[width=0.97\linewidth]{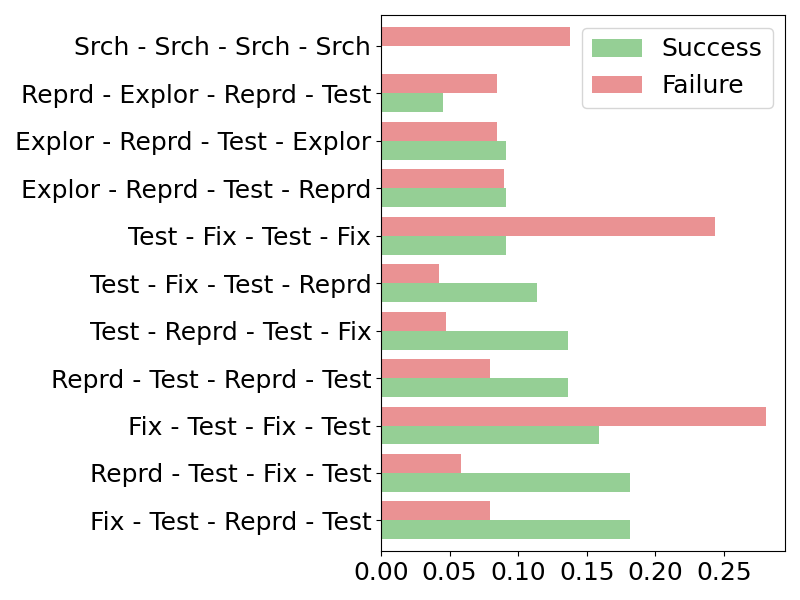}
	\caption{Top 12 sequences of action 4-grams (\oh{}).}
	\label{fig:seqopenhandsbars}
\end{figure}

\begin{figure}
	\centering
	\includegraphics[width=0.97\linewidth]{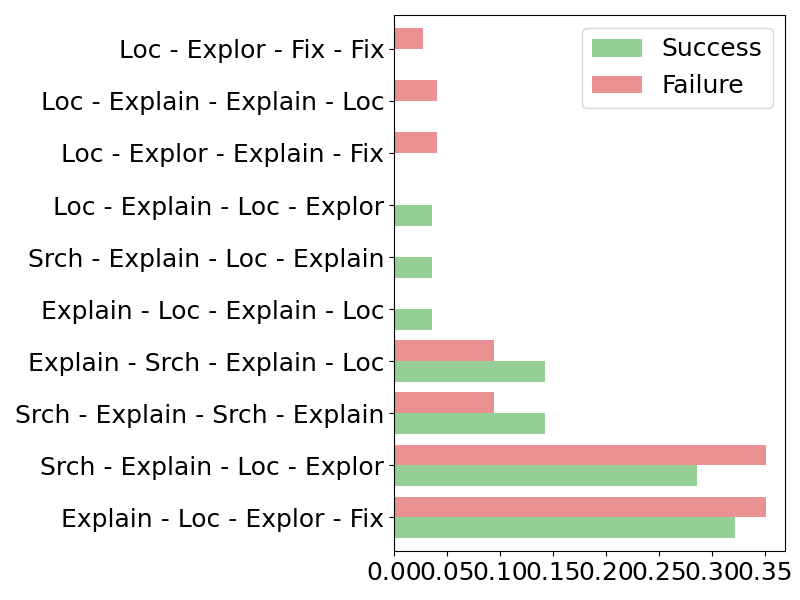}
	\caption{Top 11 sequences of action 4-grams (\acr{}).}
	\label{fig:seqautocoderoverbars}
\end{figure}

To complement the time-distributed action plots, Figures~\ref{fig:seqrepairagentbars}, \ref{fig:seqopenhandsbars}, \ref{fig:seqautocoderoverbars} presents a visualization of the most frequent 4-gram sequences executed by each agent. Each row represents a distinct sequence and its corresponding frequency in successful and unsuccessful trajectories.

In the case of \ra{}, successful 4-grams prominently feature cycles that alternate between \textit{Generate Fix} and exploratory actions, such as \textit{Explain} or \textit{Explore}, with little dominance by any single pattern, as shown in Figure~\ref{fig:seqrepairagentbars}. In contrast, unsuccessful trajectories often show many repetitive \textit{Generate Fix}--\textit{Running Tests} cycles, as shown in the last row of Figure~\ref{fig:seqrepairagentbars}.

For \oh{} (Figure~\ref{fig:seqopenhandsbars}), the most frequent successful sequences balances context gathering, patch generating, and testing (both generating and running tests). Unsuccessful trajectories, on the other hand, show more repetitive cycles of exploration and search actions (e.g., first rows in the figure), or sequences of repeated testing and fixing steps without interleaving actions. This highlights the importance balancing different actions and enforcing different debugging phases instead of continuously repeating the same actions.

For \acr{} (Figure~\ref{fig:seqautocoderoverbars}), the most common 4-grams in successful trajectories typically involve sequences with \textit{Explain} and \textit{Explore}, often followed by \textit{Generate Fix}. Notably, successful sequences tend to cluster around varied exploratory and explanatory actions.

\begin{findingsbox}{Actions Sequences}
	Analysis of frequent 4-gram action sequences shows that successful trajectories balance exploration, explanation, and validation steps, while unsuccessful ones exhibit repetitive, non-adaptive action cycles. 
\end{findingsbox}

In addition to sequence patterns linked to success, we identified debugging anti-patterns, i.e, action sequences causing incoherent results, resource waste, or unclear objectives, akin to trajectory smells described in \cite{rondon2025evaluating}. Examples include:

\textbf{1) Repeated identical actions without follow-up:} For instance, repeated search actions occur across all three agents. While some repetitions arise from unsuccessful prior searches, others neglect to act on the found results. RQ3 elaborates on this no-influence phenomenon.

\textbf{2) Repeated fix generation without testing:} As observed mainly in \oh{} and \acr{}, agents sometimes generate fixes without testing them. \ra{} mandates testing after fix generation, as it is crucial to validate that a fix works and identify any necessary patch modifications.

\textbf{3) Task termination without proper test validation:} Although \acr{} omits test validation by design, \oh{} sometimes generates reproducing tests without verifying their effectiveness (i.e, the ability of the test to reveal the bug), leading to unreliable patch validation at the end.

\begin{findingsbox}{Debugging Anti-Patterns}
We identified recurrent debugging anti-patterns that reduce effectiveness: (1) repeated identical actions without follow-up, causing wasted effort; (2) consecutive \textit{Generate Fix} steps without intermediate testing, delaying patch evaluation; and (3) premature task termination without a proper test validation. Robust agents design should include checks for anti-patterns and trajectory smells.
\end{findingsbox}

\subsection{\textbf{RQ3}: Semantic Relationships}

The semantic modeling in Sec.~\ref{sec:semantic_labeling} offers a structured lens to dissect agent decision-making.
It enables us to identify where planning fails, execution diverges, or the agent does not learn from its outputs.
Here, we provide deeper analysis to highlight tendencies and differences among agents.

\begin{table*}[t]
	\centering
	\caption{Semantic relationships between components of agent trajectories (NI = No-Influence, Ctr=Contradiction).}
	\label{tab:unholytable}
	\resizebox{\textwidth}{!}{
		\begin{tabular}{
				l
				*{2}{r}
				*{5}{r}
				*{4}{r}
				*{4}{r}
				*{3}{r}
			}
			\toprule
			\multirow{2}{*}{\textbf{Agent}}
			& \multicolumn{2}{c}{\textbf{Thought–Action}}
			& \multicolumn{5}{c}{\textbf{Thought–Thought}}
			& \multicolumn{4}{c}{\textbf{Action–Action}}
			& \multicolumn{4}{c}{\textbf{Result–Thought}}
			& \multicolumn{3}{c}{\textbf{Result–Action}} \\
			\cmidrule(lr){2-3}
			\cmidrule(lr){4-8}
			\cmidrule(lr){9-12}
			\cmidrule(lr){13-16}
			\cmidrule(lr){17-19}
			& Aln
			& Mis
			& Ctr
			& Div
			& Fol
			& Red
			& Ref
			& Div
			& Fol
			& Ref
			& Rep
			& Fol
			& MisInt
			& NI
			& Ref
			& Inf
			& NI
			& Trig \\
			\midrule
			\textbf{\ra{}}
			& 98.5 & 1.5 
			& 1.4 & 6.3 & 63.8 & 13.3 & 15.1
			& 20.6 & 50.9 & 16.2 & 12.3
			& 76.1 & 3.6 & 3.9 & 16.4
			& 25.8 & 5.3 & 68.9 \\
			- Success
			& 97.7 & 2.3 
			& 1.2 & 4.1 & 73.1 & 6.9 & 14.7
			& 23.3 & 57.9 & 12.7 & 6.1
			& 81.5 & 4.8 & 2.0 & 11.7
			& 41.2 & 1.2 & 57.6 \\
			- Failure
			& 98.7 & 1.3 
			& 1.4 & 6.8 & 61.9 & 14.7 & 15.2
			& 20.0 & 49.4 & 17.0 & 13.6
			& 74.9 & 3.4 & 4.3 & 17.4
			& 22.5 & 6.2 & 71.3 \\
			\addlinespace
			\textbf{\oh{}}
			& 98.7 & 1.3 
			& 0.0 & 14.8 & 72.8 & 1.9 & 10.5
			& 25.5 & 57.0 & 14.2 & 3.2
			& 72.1 & 3.9 & 6.3 & 17.6
			& 29.2 & 3.9 & 66.9 \\
			- Success
			& 99.5 & 0.5 
			& 0.0 & 8.6 & 74.5 & 2.0 & 14.8
			& 28.6 & 55.6 & 15.3 & 0.5
			& 73.9 & 2.3 & 6.1 & 17.7
			& 29.6 & 2.8 & 67.6 \\
			- Failure
			& 98.6 & 1.4 & 0.1 & 16.2 & 72.4 & 1.8 & 9.6
			& 24.9 & 57.4 & 14.0 & 3.8
			& 71.6 & 4.4 & 6.4 & 17.6
			& 29.1 & 4.2 & 66.7 \\
			\addlinespace
			\textbf{ACR}
			& 96.3 & 3.7 
			& 0.0 & 1.1 & 75.8 & 2.8 & 20.2
			& 24.2 & 64.6 & 4.5 & 6.7
			& 62.2 & 0.0 & 9.4 & 28.3
			& 38.8 & 12.4 & 48.9 \\
			- Success
			& 100.0 & 0.0 
			& 0.0 & 2.4 & 78.1 & 0.0 & 19.5
			& 26.8 & 68.3 & 4.8 & 0.0
			& 64.3 & 0.0 & 11.9 & 23.8
			& 43.9 & 4.9 & 51.2 \\
			- Failure
			& 95.2 & 4.8 
			& 0.0 & 0.7 & 75.2 & 3.7 & 20.4
			& 23.4 & 63.5 & 4.4 & 8.8
			& 61.6 & 0.0 & 8.7 & 29.7
			& 37.2 & 14.6 & 48.2 \\
			\bottomrule
		\end{tabular}
	}
\end{table*}

Table~\ref{tab:unholytable} presents the distribution of semantic relationships across trajectories of the three agents.
Columns represent key relational categories (e.g., Divergence, Follow-up, Refinement...).
Rows show the prevalence of each relation for both successful and unsuccessful trajectories.

\subsubsection{\textbf{Analysis of Thought-Action Relationships}}
An agent's thought is intended to guide the subsequent action and its parameters.
If the action does not align with the thought, the agent may be misled. For example, if a thought suggests running tests relevant to a bug, but the action executes unrelated tests, the agent may incorrectly conclude the issue is resolved when the tests pass while the correct tests were not executed.

Among the three agents, misalignment between thought and action is rare but closely associated with failing trajectories.
For example, \oh{} shows 0.5\% misalignment in successful and 1.4\% in failing trajectories, while \acr{} has 0\% in successful and 4.8\% in failing ones.
Although not all misalignments result in failure, even a single instance can significantly increase trajectory length or cause failure.
Furthermore, misalignment is more prevalent in failing trajectories with 2.7 misalignment counts in failing trajectories vs one misalignment in successful trajectories of OpenHands (that contain at least one misalignment).

An example of misalignment between intent and action occurs at iteration 5 during \ra{}'s execution on bug \textit{Compress\_13}.
The agent, after identifying a plausible fix, attempted refinement but instead suggested an empty fix.
This error caused confusion and extended the process from iteration 6 (where the correct patch was available) to iteration 38, increasing computational cost.
Although such misalignment is uncommon and model-dependent, incorporating mechanisms to verify that actions align with articulated thoughts is prudent
Prior work has explored this for LLMs and agents~\cite{valmeekam2023can, gou2023critic}.
Newer reasoning models (e.g., o3, deepseek, Claude 4) embed self-reflection and criticism.
Integrating these into software engineering agents, where precise translation of natural language to actions is critical, would improve robustness and efficiency in a significant fraction of trajectories.

\begin{findingsbox}{Thought-Action Alignment}
Even a single misalignment between thought and action can cause failure or increase computational cost. 
We re\-commend explicit validation that actions align with thoughts, e.g., through self-reflection, critique frameworks, or reasoning models.
\end{findingsbox}

\subsubsection{\textbf{Analysis of Thought-Thought Relationships}}

Unlike chain-of-thought reasoning, LLM-agent thoughts are interleaved with actions and their outcomes, influencing subsequent thoughts.
Table~\ref{tab:unholytable} shows that most thoughts are follow-ups or refinements of previous ones, though distinct patterns emerge per agent.
For example, \ra{} has the highest number of contradicting thoughts, which are rare in \oh{} and \acr{}.
GPT-3.5, used by \ra{}, often makes trivial mistakes.
In bug \textit{Compress\_13}, \ra{} states at iteration 5: \textit{The fix was successful, and all test cases passed.}, but contradicts this at iteration 6: \textit{The previous fixes did not work, and the bug is still present}.
Additionally, \ra{} frequently produces redundant thoughts, especially in failing trajectories, often repeating generic statements like \textit{The previous fix failed, I need to suggest a new one} without explaining failure reasons or next steps.

Divergence between thoughts is notable in \ra{} (29/40 trajectories) and \oh{} (24/40), with both positive and negative effects.
Diverging from promising paths may mislead the agent, while diverging from unproductive loops enables exploration and balances exploitation and exploration.
For example, in \oh{}'s \textit{Django-10924}, iterations 1–6 focus on understanding the implementation of \textit{FilePathField}, but iteration 7 shifts to reproducing the issue, stopping repetitive code reading in favor of testing and fixing.

\begin{findingsbox}{Relationships between Consecutive Thoughts}
Follow-ups and refinements illustrate iterative reasoning in LLM agents. Frequent contradictions and redundancies lead to confusion and resource waste. Developers should minimize contradictions and redundancies, while balancing exploitation and exploration using example trajectories or mutation-like mechanisms.
\end{findingsbox}

\subsubsection{\textbf{Analysis of Consecutive Actions Pairs}}

The actions taken by an agent should logically follow from prior actions, reflecting a coherent strategy.
Across all three agents, \emph{Follow-up} relationships dominate, indicating most actions build on their immediate predecessors.
However, the distribution of other relationship types varies among agents.
Divergence is prominent in Action-Action relationships and is often beneficial, reflecting transitions between phases (e.g., from bug localization to fix suggestion).
This is evident as action divergence is higher in successful trajectories.
Conversely, repetitive sequences and consecutive action repetition, particularly in \ra{} and \acr{}, strongly correlate with failing trajectories.
Repetition wastes iterations and may induce unproductive loops, as also observed in RQ2. As shown in the table, action repetition is several times higher in failing trajectories.
Repetition can be detected using sequence analysis methods such as sliding window matching or n-gram frequency analysis.
Detected repetitions can then be addressed by diversification strategies that encourage exploration of alternative actions or transitions.

\begin{findingsbox}{Actionable Insights on Action Sequence Dynamics}
Repetition can be mitigated using sequence analysis methods, 
applying diversification strategies, including alternative actions or exploration steps to break repetitive cycles and cover new paths.
\end{findingsbox}

\subsubsection{\textbf{Influence of Results on Thoughts and Actions}}
\paragraph{Analysis of Result–Thought Relationships}
Ideally, agents should interpret results accurately, using informative signals to guide their reasoning, correct errors, or refine strategies.
However, Table~\ref{tab:unholytable} shows both strengths and limitations in how agents process and respond to results.

Across all agents, the most common relationship is follow-up, where a thought logically builds on the received result, such as responding to a failed test by suggesting a new patch.
Refinement relationships are also frequent, indicating efforts to improve the solution based on results.
Misinterpretations and no influence cases are less common but occur in both failing and successful instances. For example, in \ra{}’s run on bug \textit{Cli\_11} at iteration 6, the agent misinterprets the result indicating that the last fix was empty and instead of suggesting a non-empty fix, it moves on to analyzing the code with the intention to change \textit{empty} function arguments.

\paragraph{Analysis of Result–Action Relationships}
The outcome of an action should shape the agent's next action.
In successful runs, we observe a high proportion of triggering actions, where the result directly prompts a relevant next step.
Informative actions are also common, as agents refine their actions based on information drawn from previous results.

However, failures often correlate with a higher proportion of no-influence actions (1.2\% vs.\ 6.2\% for \ra{} and 4.9\% vs 14.6\% for \acr{}), where the agent’s next move does not reflect the information content of the preceding result. These observations suggest that effective LLM agents not only need to interpret results accurately at the reasoning level but also translate those insights into appropriately responsive action sequences.

\begin{findingsbox}{Result Driven Reasoning and Actions}  
Improving LLM-agents requires mechanisms that enhance result sensitivity, ensuring results consistently guide both reasoning and action decisions towards suggesting an effective next actions.
\end{findingsbox}

\section{Limitations and Threats to Validity}  

Our study offers insights into LLM-based agents in software engineering tasks but faces limitations, notably the potential subjectivity in our annotation and classification process despite partial inter-annotator agreement checks. We evaluate three bug-fixing agents which, despite their common use, do not fully capture all software development tasks or real-world deployment scenarios. The agents rely on distinct models, ensuring diversity but complicating fair comparisons. The generalization of our results to other software engineering agents or tasks, such as code generation or testing, remains open for study.

Furthermore, the sampled trajectories used for in-depth analysis may not capture the full variability in agent behavior. Although we apply a random sampling strategy to ensure diversity, rare but important patterns could be overlooked. Future studies could address this limitation by scaling up the dataset and using automated pattern detection techniques.

Our correlation and sequence mining analyses reveal trends and associations without establishing causality between behaviors and task success. Despite this limitation, our study provides a structured approach to analyzing LLM-based agent behavior and offers valuable insights for future research and autonomous software engineering tool development.

\section{Related Work}

\paragraph{Deep Learning and LLMs in Software Engineering}

Neural reasoning, particularly with LLMs, has attracted significant attention in software engineering research~\cite{NeuralSoftwareAnalysis}.
Early work focused on neural code completion~\cite{Chen2021,ziegler2022productivity,arXiv2024_De-Hallucinator,Barke2023}.
Recent studies extend LLMs to tasks such as program repair~\cite{Jiang2023,Xia2024a,hidvegi2024cigar}, issue solving~\cite{Xia2024}, test case generation~\cite{Lemieux2023,Ryan2024,Yuan2024}, test oracle generation~\cite{Hayet2024a,Hossain2025}, and fuzz testing~\cite{icse2024-Fuzz4All}.
Unlike the agentic approaches studied in this paper, these methods prompt LLMs with pre-defined algorithms.

\paragraph{LLM Agents}

Recently, LLM agents have emerged as a new paradigm in software engineering, where LLMs autonomously solve complex tasks by reasoning and interacting with external tools~\cite{AgenticAISE2025}.
A prominent application is automated bug and issue fixing, with numerous agents proposed, including
RepairAgent~\cite{icse2025-RepairAgent},
AutoCodeRover~\cite{zhang2024autocoderover},
OpenHands~\cite{wang2024openhands},
SWE-Agent~\cite{Yang2024a},
Magis~\cite{Tao2024},
AgentCoder~\cite{Huang2024},
MarsCode Agent~\cite{Liu2024a},
FixAgent~\cite{Lee2024}, and 
Passerine~\cite{Rondon2025}.
Motivated by this task's relevance, our work studies three representative agents.
Other agents focus on writing tests to reproduce issues from natural language~\cite{Muendler2024,Ahmed2024,Issue2Test_arXiv2025,Cheng2025},
automating project setup~\cite{issta2025_ExecutionAgent,Milliken2024,hu2025llm},
debugging computational notebooks~\cite{Grotov2024}, and
analyzing root causes of failures from logs~\cite{Roy2024}.
To support progress, several benchmarks have been introduced, such as SWE-bench~\cite{Jimenez2023}, SWE-bench+~\cite{Aleithan2024}, Env\-Bench~\cite{Eliseeva2025}, and automated benchmark generation~\cite{Yang2025}.
For a comprehensive overview, we refer to recent surveys~\cite{wang2024agents,jin2024llms}.
Rather than proposing a new agent, our work systematically analyzes existing agents to inform future agent design.

\paragraph{AI Interpretability and Explainability}

Our work relates to efforts on interpreting and explaining AI models.
These include techniques that identify relevant nodes and edges in graph-based neural models for specific predictions, e.g., in NLP~\cite{ying2019gnnexplainer,Schlichtkrull2021} or vulnerability detection~\cite{Li2021c}.
Other work explains predictions via counterfactual reasoning~\cite{Cito2022} or by learning predicates that characterize mispredicted inputs~\cite{Cito2021}.
Yeo et al.\ analyze chain-of-thought reasoning, focusing on the impact of training and feedback~\cite{yeo2025demystifying}.
Surveys provide overviews of explanation techniques for black-box models~\cite{Guidotti2019} and LLMs~\cite{survey_explain_llm}.
Recent work also targets LLM agents, e.g., by proposing progress metrics for trajectory analysis~\cite{chang2024agentboard}, visualizing agent behavior in multi-agent systems~\cite{lu2024agentlens}, or offering debugger-like interfaces~\cite{Epperson2025}.
To our knowledge, our work is the first to systematically study the reasoning and decision-making processes of multiple LLM-based agents in software engineering.

\paragraph{Automatically Improving LLM Agents}
To improve LLM agents, one approach applies reinforcement learning based on past trajectories and their outcomes~\cite{Gupta2024,song2024trial,deng2024novice}.
Another direction fine-tunes LLMs using previous agent trajectories to enhance performance on specific tasks or optimize multi-agent communication~\cite{song2024agentbank,chen2024optima}.
A third line introduces additional agents that interpret or provide feedback on the main agent's actions~\cite{Antoniades2024}.
While our work also aims to improve LLM agents, we focus on deriving insights through systematic analysis of existing agents and their tool interactions, rather than fully automating the improvement process.

\section{Conclusion}  

This study systematically analyzes LLM-based autonomous agents in software engineering, focusing on decision-making, reasoning consistency, and trajectory dynamics. Using statistical analysis, semantic relation modeling, open coding, and sequence pattern mining, we identify key behavioral trends. Successful trajectories balance information gathering, hypothesis testing, and fix validation, while unsuccessful ones exhibit redundant exploration or premature fixes. Categorizing actions and sequences reveals strengths and weaknesses of agents, guiding improvements in efficiency, reliability, and reasoning. Future work could include expanding agent and benchmark coverage, refining classification methods, and developing automatic failure detection and mitigation techniques.

\section*{Data availability}
Our code and data are  available at:

 \url{https://github.com/sola-st/llm-agents-study} 

\section*{Acknowledgments}

This work was supported by the European Research Council (ERC, grant agreements 851895 and 101155832) and by the German Research Foundation within the DeMoCo project.

\bibliographystyle{IEEEtran}    %
\bibliography{references,referencesMP}

\begin{thebibliography}{10}
\providecommand{\url}[1]{#1}
\csname url@samestyle\endcsname
\providecommand{\newblock}{\relax}
\providecommand{\bibinfo}[2]{#2}
\providecommand{\BIBentrySTDinterwordspacing}{\spaceskip=0pt\relax}
\providecommand{\BIBentryALTinterwordstretchfactor}{4}
\providecommand{\BIBentryALTinterwordspacing}{\spaceskip=\fontdimen2\font plus
\BIBentryALTinterwordstretchfactor\fontdimen3\font minus
  \fontdimen4\font\relax}
\providecommand{\BIBforeignlanguage}[2]{{%
\expandafter\ifx\csname l@#1\endcsname\relax
\typeout{** WARNING: IEEEtran.bst: No hyphenation pattern has been}%
\typeout{** loaded for the language `#1'. Using the pattern for}%
\typeout{** the default language instead.}%
\else
\language=\csname l@#1\endcsname
\fi
#2}}
\providecommand{\BIBdecl}{\relax}
\BIBdecl

\bibitem{Chen2021}
\BIBentryALTinterwordspacing
M.~Chen, J.~Tworek, H.~Jun, Q.~Yuan, H.~P. de~Oliveira~Pinto, J.~Kaplan,
  H.~Edwards, Y.~Burda, N.~Joseph, G.~Brockman, A.~Ray, R.~Puri, G.~Krueger,
  M.~Petrov, H.~Khlaaf, G.~Sastry, P.~Mishkin, B.~Chan, S.~Gray, N.~Ryder,
  M.~Pavlov, A.~Power, L.~Kaiser, M.~Bavarian, C.~Winter, P.~Tillet, F.~P.
  Such, D.~Cummings, M.~Plappert, F.~Chantzis, E.~Barnes, A.~Herbert{-}Voss,
  W.~H. Guss, A.~Nichol, A.~Paino, N.~Tezak, J.~Tang, I.~Babuschkin, S.~Balaji,
  S.~Jain, W.~Saunders, C.~Hesse, A.~N. Carr, J.~Leike, J.~Achiam, V.~Misra,
  E.~Morikawa, A.~Radford, M.~Knight, M.~Brundage, M.~Murati, K.~Mayer,
  P.~Welinder, B.~McGrew, D.~Amodei, S.~McCandlish, I.~Sutskever, and
  W.~Zaremba, ``Evaluating large language models trained on code,''
  \emph{CoRR}, vol. abs/2107.03374, 2021. [Online]. Available:
  \url{https://arxiv.org/abs/2107.03374}
\BIBentrySTDinterwordspacing

\bibitem{ziegler2022productivity}
A.~Ziegler, E.~Kalliamvakou, X.~A. Li, A.~Rice, D.~Rifkin, S.~Simister,
  G.~Sittampalam, and E.~Aftandilian, ``Productivity assessment of neural code
  completion,'' in \emph{Proceedings of the 6th ACM SIGPLAN International
  Symposium on Machine Programming}, 2022, pp. 21--29.

\bibitem{arXiv2024_De-Hallucinator}
\BIBentryALTinterwordspacing
A.~Eghbali and M.~Pradel, ``De-hallucinator: Iterative grounding for llm-based
  code completion,'' \emph{CoRR}, vol. abs/2401.01701, 2024. [Online].
  Available: \url{https://doi.org/10.48550/arXiv.2401.01701}
\BIBentrySTDinterwordspacing

\bibitem{Barke2023}
\BIBentryALTinterwordspacing
S.~Barke, M.~B. James, and N.~Polikarpova, ``Grounded copilot: How programmers
  interact with code-generating models,'' \emph{Proc. {ACM} Program. Lang.},
  vol.~7, no. {OOPSLA1}, pp. 85--111, 2023. [Online]. Available:
  \url{https://doi.org/10.1145/3586030}
\BIBentrySTDinterwordspacing

\bibitem{Jiang2023}
\BIBentryALTinterwordspacing
N.~Jiang, K.~Liu, T.~Lutellier, and L.~Tan, ``Impact of code language models on
  automated program repair,'' in \emph{ICSE}, 2023, pp. 1430--1442. [Online].
  Available: \url{https://doi.org/10.1109/ICSE48619.2023.00125}
\BIBentrySTDinterwordspacing

\bibitem{Xia2024a}
\BIBentryALTinterwordspacing
C.~S. Xia and L.~Zhang, ``Automated program repair via conversation: Fixing 162
  out of 337 bugs for {\textdollar}0.42 each using chatgpt,'' in
  \emph{Proceedings of the 33rd {ACM} {SIGSOFT} International Symposium on
  Software Testing and Analysis, {ISSTA} 2024, Vienna, Austria, September
  16-20, 2024}, M.~Christakis and M.~Pradel, Eds.\hskip 1em plus 0.5em minus
  0.4em\relax {ACM}, 2024, pp. 819--831. [Online]. Available:
  \url{https://doi.org/10.1145/3650212.3680323}
\BIBentrySTDinterwordspacing

\bibitem{Lemieux2023}
C.~Lemieux, J.~P. Inala, S.~K. Lahiri, and S.~Sen, ``Codamosa: Escaping
  coverage plateaus in test generation with pre-trained large language
  models,'' in \emph{45th International Conference on Software Engineering,
  ser. ICSE}, 2023.

\bibitem{Ryan2024}
G.~Ryan, S.~Jain, M.~Shang, S.~Wang, X.~Ma, M.~K. Ramanathan, and B.~Ray,
  ``Code-aware prompting: A study of coverage guided test generation in
  regression setting using llm,'' in \emph{FSE}, 2024.

\bibitem{Yuan2024}
\BIBentryALTinterwordspacing
Z.~Yuan, M.~Liu, S.~Ding, K.~Wang, Y.~Chen, X.~Peng, and Y.~Lou, ``Evaluating
  and improving chatgpt for unit test generation,'' \emph{Proc. {ACM} Softw.
  Eng.}, vol.~1, no. {FSE}, pp. 1703--1726, 2024. [Online]. Available:
  \url{https://doi.org/10.1145/3660783}
\BIBentrySTDinterwordspacing

\bibitem{Hayet2024a}
I.~Hayet, A.~Scott, and M.~d’Amorim, ``Chatassert: Llm-based test oracle
  generation with external tools assistance,'' \emph{IEEE Transactions on
  Software Engineering}, pp. 1--15, 2024.

\bibitem{Hossain2025}
S.~B. Hossain and M.~B. Dwyer, ``Togll: Correct and strong test oracle
  generation with llms,'' in \emph{ICSE}, 2025.

\bibitem{icse2024-Fuzz4All}
\BIBentryALTinterwordspacing
C.~S. Xia, M.~Paltenghi, J.~L. Tian, M.~Pradel, and L.~Zhang, ``Fuzz4all:
  Universal fuzzing with large language models,'' in \emph{Proceedings of the
  46th {IEEE/ACM} International Conference on Software Engineering, {ICSE}
  2024, Lisbon, Portugal, April 14-20, 2024}.\hskip 1em plus 0.5em minus
  0.4em\relax {ACM}, 2024, pp. 126:1--126:13. [Online]. Available:
  \url{https://doi.org/10.1145/3597503.3639121}
\BIBentrySTDinterwordspacing

\bibitem{AgenticAISE2025}
A.~Roychoudhury, C.~Pasareanu, M.~Pradel, and B.~Ray, ``Agentic ai software
  engineer: Programming with trust,'' \emph{arXiv preprint arXiv:2502.13767},
  2025.

\bibitem{icse2025-RepairAgent}
I.~Bouzenia, P.~Devanbu, and M.~Pradel, ``{RepairAgent}: An autonomous,
  {LLM}-based agent for program repair,'' in \emph{International Conference on
  Software Engineering (ICSE)}, 2025.

\bibitem{zhang2024autocoderover}
Y.~Zhang, H.~Ruan, Z.~Fan, and A.~Roychoudhury, ``Autocoderover: Autonomous
  program improvement,'' in \emph{Proceedings of the 33rd ACM SIGSOFT
  International Symposium on Software Testing and Analysis}, 2024, pp.
  1592--1604.

\bibitem{wang2024openhands}
X.~Wang, B.~Li, Y.~Song, F.~F. Xu, X.~Tang, M.~Zhuge, J.~Pan, Y.~Song, B.~Li,
  J.~Singh \emph{et~al.}, ``Openhands: An open platform for ai software
  developers as generalist agents,'' in \emph{The Thirteenth International
  Conference on Learning Representations}, 2024.

\bibitem{Yang2024a}
\BIBentryALTinterwordspacing
J.~Yang, C.~E. Jimenez, A.~Wettig, K.~Lieret, S.~Yao, K.~Narasimhan, and
  O.~Press, ``Swe-agent: Agent-computer interfaces enable automated software
  engineering,'' in \emph{Advances in Neural Information Processing Systems 38:
  Annual Conference on Neural Information Processing Systems 2024, NeurIPS
  2024, Vancouver, BC, Canada, December 10 - 15, 2024}, A.~Globersons,
  L.~Mackey, D.~Belgrave, A.~Fan, U.~Paquet, J.~M. Tomczak, and C.~Zhang, Eds.,
  2024. [Online]. Available:
  \url{http://papers.nips.cc/paper\_files/paper/2024/hash/5a7c947568c1b1328ccc5230172e1e7c-Abstract-Conference.html}
\BIBentrySTDinterwordspacing

\bibitem{Muendler2024}
\BIBentryALTinterwordspacing
N.~Mündler, M.~N. Müller, J.~He, and M.~Vechev, ``Code agents are state of
  the art software testers,'' 2024. [Online]. Available:
  \url{https://arxiv.org/abs/2406.12952}
\BIBentrySTDinterwordspacing

\bibitem{Ahmed2024}
\BIBentryALTinterwordspacing
T.~Ahmed, M.~Hirzel, R.~Pan, A.~Shinnar, and S.~Sinha, ``Tdd-bench verified:
  Can llms generate tests for issues before they get resolved?'' 2024.
  [Online]. Available: \url{https://arxiv.org/abs/2412.02883}
\BIBentrySTDinterwordspacing

\bibitem{Issue2Test_arXiv2025}
N.~Nashid, I.~Bouzenia, M.~Pradel, and A.~Mesbah, ``Issue2test: Generating
  reproducing test cases from issue reports,'' \emph{arXiv preprint
  arXiv:2503.16320}, 2025.

\bibitem{hu2025llm}
R.~Hu, C.~Peng, X.~Wang, and C.~Gao, ``An llm-based agent for reliable docker
  environment configuration,'' \emph{arXiv preprint arXiv:2502.13681}, 2025.

\bibitem{issta2025_ExecutionAgent}
I.~Bouzenia and M.~Pradel, ``You name it, {I} run it: An {LLM} agent to execute
  tests of arbitrary projects,'' in \emph{ISSTA}, 2025.

\bibitem{Guidotti2019}
\BIBentryALTinterwordspacing
R.~Guidotti, A.~Monreale, S.~Ruggieri, F.~Turini, F.~Giannotti, and
  D.~Pedreschi, ``A survey of methods for explaining black box models,''
  \emph{{ACM} Comput. Surv.}, vol.~51, no.~5, pp. 93:1--93:42, 2019. [Online].
  Available: \url{https://doi.org/10.1145/3236009}
\BIBentrySTDinterwordspacing

\bibitem{survey_explain_llm}
\BIBentryALTinterwordspacing
H.~Zhao, H.~Chen, F.~Yang, N.~Liu, H.~Deng, H.~Cai, S.~Wang, D.~Yin, and M.~Du,
  ``Explainability for large language models: A survey,'' \emph{ACM Trans.
  Intell. Syst. Technol.}, vol.~15, no.~2, Feb. 2024. [Online]. Available:
  \url{https://doi.org/10.1145/3639372}
\BIBentrySTDinterwordspacing

\bibitem{Cito2022}
J.~Cito, I.~Dillig, V.~Murali, and S.~Chandra, ``Counterfactual explanations
  for models of code,'' in \emph{ICSE-SEIP}, 2022.

\bibitem{ying2019gnnexplainer}
Z.~Ying, D.~Bourgeois, J.~You, M.~Zitnik, and J.~Leskovec, ``Gnnexplainer:
  Generating explanations for graph neural networks,'' \emph{Advances in neural
  information processing systems}, vol.~32, 2019.

\bibitem{lu2024agentlens}
J.~Lu, B.~Pan, J.~Chen, Y.~Feng, J.~Hu, Y.~Peng, and W.~Chen, ``Agentlens:
  Visual analysis for agent behaviors in llm-based autonomous systems,''
  \emph{IEEE Transactions on Visualization and Computer Graphics}, 2024.

\bibitem{Epperson2025}
W.~Epperson, G.~Bansal, V.~Dibia, A.~Fourney, J.~Gerrits, E.~Zhu, and
  S.~Amershi, ``Interactive debugging and steering of multi-agent ai systems,''
  \emph{arXiv preprint arXiv:2503.02068}, 2025.

\bibitem{Gupta2024}
\BIBentryALTinterwordspacing
P.~Gupta, S.~Kirtania, A.~Singha, S.~Gulwani, A.~Radhakrishna, G.~Soares, and
  S.~Shi, ``Metareflection: Learning instructions for language agents using
  past reflections,'' in \emph{Proceedings of the 2024 Conference on Empirical
  Methods in Natural Language Processing, {EMNLP} 2024, Miami, FL, USA,
  November 12-16, 2024}, Y.~Al{-}Onaizan, M.~Bansal, and Y.~Chen, Eds.\hskip
  1em plus 0.5em minus 0.4em\relax Association for Computational Linguistics,
  2024, pp. 8369--8385. [Online]. Available:
  \url{https://aclanthology.org/2024.emnlp-main.477}
\BIBentrySTDinterwordspacing

\bibitem{song2024trial}
Y.~Song, D.~Yin, X.~Yue, J.~Huang, S.~Li, and B.~Y. Lin, ``Trial and error:
  Exploration-based trajectory optimization for llm agents,'' \emph{arXiv
  preprint arXiv:2403.02502}, 2024.

\bibitem{deng2024novice}
Z.~Deng, Z.~Dou, Y.~Zhu, J.-R. Wen, R.~Xiong, M.~Wang, and W.~Chen, ``From
  novice to expert: Llm agent policy optimization via step-wise reinforcement
  learning,'' \emph{arXiv preprint arXiv:2411.03817}, 2024.

\bibitem{Just2014}
R.~Just, D.~Jalali, and M.~D. Ernst, ``Defects4j: a database of existing faults
  to enable controlled testing studies for java programs,'' in \emph{ISSTA},
  2014, pp. 437--440.

\bibitem{Jimenez2023}
C.~E. Jimenez, J.~Yang, A.~Wettig, S.~Yao, K.~Pei, O.~Press, and K.~Narasimhan,
  ``Swe-bench: Can language models resolve real-world github issues?'' 2023.

\bibitem{zeller2009programs}
A.~Zeller, \emph{Why programs fail: a guide to systematic debugging}.\hskip 1em
  plus 0.5em minus 0.4em\relax Morgan Kaufmann, 2009.

\bibitem{corbin1990basics}
J.~Corbin \emph{et~al.}, ``Basics of qualitative research grounded theory
  procedures and techniques,'' 1990.

\bibitem{ruan2024specrover}
H.~Ruan, Y.~Zhang, and A.~Roychoudhury, ``Specrover: Code intent extraction via
  llms,'' \emph{arXiv preprint arXiv:2408.02232}, 2024.

\bibitem{rondon2025evaluating}
P.~Rondon, R.~Wei, J.~Cambronero, J.~Cito, A.~Sun, S.~Sanyam, M.~Tufano, and
  S.~Chandra, ``Evaluating agent-based program repair at google,'' \emph{arXiv
  preprint arXiv:2501.07531}, 2025.

\bibitem{valmeekam2023can}
K.~Valmeekam, M.~Marquez, and S.~Kambhampati, ``Can large language models
  really improve by self-critiquing their own plans?'' \emph{arXiv preprint
  arXiv:2310.08118}, 2023.

\bibitem{gou2023critic}
Z.~Gou, Z.~Shao, Y.~Gong, Y.~Shen, Y.~Yang, N.~Duan, and W.~Chen, ``Critic:
  Large language models can self-correct with tool-interactive critiquing,''
  \emph{arXiv preprint arXiv:2305.11738}, 2023.

\bibitem{NeuralSoftwareAnalysis}
\BIBentryALTinterwordspacing
M.~Pradel and S.~Chandra, ``Neural software analysis,'' \emph{Commun. {ACM}},
  vol.~65, no.~1, pp. 86--96, 2022. [Online]. Available:
  \url{https://doi.org/10.1145/3460348}
\BIBentrySTDinterwordspacing

\bibitem{hidvegi2024cigar}
D.~Hidv{\'e}gi, K.~Etemadi, S.~Bobadilla, and M.~Monperrus, ``Cigar:
  Cost-efficient program repair with llms,'' \emph{arXiv preprint
  arXiv:2402.06598}, 2024.

\bibitem{Xia2024}
\BIBentryALTinterwordspacing
C.~S. Xia, Y.~Deng, S.~Dunn, and L.~Zhang, ``Agentless: Demystifying llm-based
  software engineering agents,'' 2024. [Online]. Available:
  \url{https://arxiv.org/abs/2407.01489}
\BIBentrySTDinterwordspacing

\bibitem{Tao2024}
W.~Tao, Y.~Zhou, W.~Zhang, and Y.~Cheng, ``Magis: Llm-based multi-agent
  framework for github issue resolution,'' \emph{arXiv preprint
  arXiv:2403.17927}, 2024.

\bibitem{Huang2024}
D.~Huang, Q.~Bu, J.~M. Zhang, M.~Luck, and H.~Cui, ``Agentcoder:
  Multi-agent-based code generation with iterative testing and optimisation,''
  2024.

\bibitem{Liu2024a}
Y.~Liu, P.~Gao, X.~Wang, C.~Peng, and Z.~Zhang, ``Marscode agent: Ai-native
  automated bug fixing,'' \emph{arXiv preprint arXiv:2409.00899}, 2024.

\bibitem{Lee2024}
\BIBentryALTinterwordspacing
C.~Lee, C.~S. Xia, L.~Yang, J.~tse Huang, Z.~Zhu, L.~Zhang, and M.~R. Lyu, ``A
  unified debugging approach via llm-based multi-agent synergy,'' 2024.
  [Online]. Available: \url{https://arxiv.org/abs/2404.17153}
\BIBentrySTDinterwordspacing

\bibitem{Rondon2025}
\BIBentryALTinterwordspacing
P.~Rondon, R.~Wei, J.~Cambronero, J.~Cito, A.~Sun, S.~Sanyam, M.~Tufano, and
  S.~Chandra, ``Evaluating agent-based program repair at google,'' 2025.
  [Online]. Available: \url{https://arxiv.org/abs/2501.07531}
\BIBentrySTDinterwordspacing

\bibitem{Cheng2025}
R.~Cheng, M.~Tufano, J.~Cito, J.~Cambronero, P.~Rondon, R.~Wei, A.~Sun, and
  S.~Chandra, ``Agentic bug reproduction for effective automated program repair
  at google,'' \emph{arXiv preprint arXiv:2502.01821}, 2025.

\bibitem{Milliken2024}
\BIBentryALTinterwordspacing
L.~Milliken, S.~Kang, and S.~Yoo, ``Beyond pip install: Evaluating llm agents
  for the automated installation of python projects,'' 2024. [Online].
  Available: \url{https://arxiv.org/abs/2412.06294}
\BIBentrySTDinterwordspacing

\bibitem{Grotov2024}
K.~Grotov, A.~Borzilov, M.~Krivobok, T.~Bryksin, and Y.~Zharov, ``Debug
  smarter, not harder: Ai agents for error resolution in computational
  notebooks,'' \emph{arXiv preprint arXiv:2410.14393}, 2024.

\bibitem{Roy2024}
\BIBentryALTinterwordspacing
D.~Roy, X.~Zhang, R.~Bhave, C.~Bansal, P.~H.~B. Las{-}Casas, R.~Fonseca, and
  S.~Rajmohan, ``Exploring llm-based agents for root cause analysis,'' in
  \emph{Companion Proceedings of the 32nd {ACM} International Conference on the
  Foundations of Software Engineering, {FSE} 2024, Porto de Galinhas, Brazil,
  July 15-19, 2024}, M.~d'Amorim, Ed.\hskip 1em plus 0.5em minus 0.4em\relax
  {ACM}, 2024, pp. 208--219. [Online]. Available:
  \url{https://doi.org/10.1145/3663529.3663841}
\BIBentrySTDinterwordspacing

\bibitem{Aleithan2024}
\BIBentryALTinterwordspacing
R.~Aleithan, H.~Xue, M.~M. Mohajer, E.~Nnorom, G.~Uddin, and S.~Wang,
  ``Swe-bench+: Enhanced coding benchmark for llms,'' \emph{CoRR}, vol.
  abs/2410.06992, 2024. [Online]. Available:
  \url{https://doi.org/10.48550/arXiv.2410.06992}
\BIBentrySTDinterwordspacing

\bibitem{Eliseeva2025}
\BIBentryALTinterwordspacing
A.~Eliseeva, A.~Kovrigin, I.~Kholkin, E.~Bogomolov, and Y.~Zharov, ``Envbench:
  A benchmark for automated environment setup,'' 2025. [Online]. Available:
  \url{https://arxiv.org/abs/2503.14443}
\BIBentrySTDinterwordspacing

\bibitem{Yang2025}
J.~Yang, K.~Leret, C.~E. Jimenez, A.~Wettig, K.~Khandpur, Y.~Zhang, B.~Hui,
  O.~Press, L.~Schmidt, and D.~Yang, ``Swe-smith: Scaling data for software
  engineering agents,'' \emph{arXiv preprint arXiv:2504.21798}, 2025.

\bibitem{wang2024agents}
Y.~Wang, W.~Zhong, Y.~Huang, E.~Shi, M.~Yang, J.~Chen, H.~Li, Y.~Ma, Q.~Wang,
  and Z.~Zheng, ``Agents in software engineering: Survey, landscape, and
  vision,'' \emph{arXiv preprint arXiv:2409.09030}, 2024.

\bibitem{jin2024llms}
H.~Jin, L.~Huang, H.~Cai, J.~Yan, B.~Li, and H.~Chen, ``From llms to llm-based
  agents for software engineering: A survey of current, challenges and
  future,'' \emph{arXiv preprint arXiv:2408.02479}, 2024.

\bibitem{Schlichtkrull2021}
\BIBentryALTinterwordspacing
M.~S. Schlichtkrull, N.~D. Cao, and I.~Titov, ``Interpreting graph neural
  networks for {NLP} with differentiable edge masking,'' in \emph{9th
  International Conference on Learning Representations, {ICLR} 2021, Virtual
  Event, Austria, May 3-7, 2021}.\hskip 1em plus 0.5em minus 0.4em\relax
  OpenReview.net, 2021. [Online]. Available:
  \url{https://openreview.net/forum?id=WznmQa42ZAx}
\BIBentrySTDinterwordspacing

\bibitem{Li2021c}
\BIBentryALTinterwordspacing
Y.~Li, S.~Wang, and T.~N. Nguyen, ``Vulnerability detection with fine-grained
  interpretations,'' in \emph{{ESEC/FSE} '21: 29th {ACM} Joint European
  Software Engineering Conference and Symposium on the Foundations of Software
  Engineering, Athens, Greece, August 23-28, 2021}, D.~Spinellis, G.~Gousios,
  M.~Chechik, and M.~D. Penta, Eds.\hskip 1em plus 0.5em minus 0.4em\relax
  {ACM}, 2021, pp. 292--303. [Online]. Available:
  \url{https://doi.org/10.1145/3468264.3468597}
\BIBentrySTDinterwordspacing

\bibitem{Cito2021}
J.~Cito, I.~Dillig, S.~Kim, Vijayaraghavan, and M.~S. Chandra, ``Explaining
  mispredictions of machine learning models using rule induction,'' in
  \emph{ESEC/FSE}, 2021.

\bibitem{yeo2025demystifying}
E.~Yeo, Y.~Tong, M.~Niu, G.~Neubig, and X.~Yue, ``Demystifying long
  chain-of-thought reasoning in llms,'' \emph{arXiv preprint arXiv:2502.03373},
  2025.

\bibitem{chang2024agentboard}
M.~Chang, J.~Zhang, Z.~Zhu, C.~Yang, Y.~Yang, Y.~Jin, Z.~Lan, L.~Kong, and
  J.~He, ``Agentboard: An analytical evaluation board of multi-turn llm
  agents,'' \emph{Advances in Neural Information Processing Systems}, vol.~37,
  pp. 74\,325--74\,362, 2024.

\bibitem{song2024agentbank}
Y.~Song, W.~Xiong, X.~Zhao, D.~Zhu, W.~Wu, K.~Wang, C.~Li, W.~Peng, and S.~Li,
  ``Agentbank: Towards generalized llm agents via fine-tuning on 50000+
  interaction trajectories,'' \emph{arXiv preprint arXiv:2410.07706}, 2024.

\bibitem{chen2024optima}
W.~Chen, J.~Yuan, C.~Qian, C.~Yang, Z.~Liu, and M.~Sun, ``Optima: Optimizing
  effectiveness and efficiency for llm-based multi-agent system,'' \emph{arXiv
  preprint arXiv:2410.08115}, 2024.

\bibitem{Antoniades2024}
\BIBentryALTinterwordspacing
A.~Antoniades, A.~Örwall, K.~Zhang, Y.~Xie, A.~Goyal, and W.~Wang,
  ``Swe-search: Enhancing software agents with monte carlo tree search and
  iterative refinement,'' 2024. [Online]. Available:
  \url{https://arxiv.org/abs/2410.20285}
\BIBentrySTDinterwordspacing

\end{thebibliography}
\end{document}